\documentstyle[11pt,paspconf]{article}
\input epsf
\def\sun {$_{\scriptscriptstyle \odot}$}
\def\ltaprx {\lower .1ex\hbox{\rlap{\raise .6ex\hbox{\hskip .3ex
	{\ifmmode{\scriptscriptstyle <}\else
		{$\scriptscriptstyle <$}\fi}}}
	\kern -.4ex{\ifmmode{\scriptscriptstyle \sim}\else
		{$\scriptscriptstyle\sim$}\fi}}}

\begin{document}
\title{SN 1987A - Presupernova Evolution and the Progenitor Star}
\author{S. E. Woosley}
\affil{Board of Studies in Astronomy and Astrophysics, University of
California, Santa Cruz, CA 95064}
\author{Alexander Heger}
\affil{Max Planck Institut f\"ur Astrophysik, D--85740 Garching, Germany}
\author{Thomas A. Weaver}
\affil{General Studies Group, Physics Department, Lawrence
Livermore National Laboratory, Livermore, CA 94550}
\author{Norbert Langer}
\affil{Institut f\"ur Theoretische Physik und Astrophysik, Universit\"at
Potsdam, D--14415 Potsdam, Germany}

\begin{abstract}
Ten years later, astronomers are still puzzled by the stellar
evolution that produced SN 1987A --- a blue supergiant.  In single
star models, the new OPAL opacities make blue solutions more difficult
to achieve, though still possible for certain choices of convection
physics.  We also consider rotation, which has the desirable effect
of producing large surface enhancements of nitrogen and helium, but
the undesirable effect of increasing the helium-core mass at the
expense of the envelope.  The latter makes blue solutions more
difficult.  Still, we seek a model that occurs with high probability
in the LMC and for which the time-scale for making the last transition
from red to blue, $\sim$20$\,$000 years, has a physical interpretation
--- the Kelvin-Helmholtz time of the helium core.  Single star models
satisfy both criteria and might yet prove to be the correct
explanation for Sk -69 202, provided new rotational or convection
physics can simultaneously give a blue star and explain the ring
structure.  Some speculations on how this might be achieved are
presented and some aspects of binary models briefly discussed.
\end{abstract}


\section{Introduction}

Following the explosion of SN 1987A, a large variety of models were
proposed that sought to explain both the light curve and the
identification of the presupernova star, Sk -69 202, a blue
supergiant.  Many of these fell by the wayside as more data were
obtained and detailed models constructed.  The idea of mass loss as the
explanation of a blue star (e.g., Maeder 1987; Chevalier \& Fransson
1987; Shigeyama et al. 1987) failed to explain the long delay before
the second light peak and the slow velocity of the ejecta.  SN 1987A,
unlike SN 1993J, was not a Type IIb supernova.  The suggested low
metallicity of the LMC offered an explanation of why many blue
supernova progenitors might exist (Shklovskii 1987; Truran \& Weiss
1987; Arnett 1987; Hillebrandt et al. 1987), but, by itself, failed to
explain the existence of a nitrogen rich circumstellar shell
surrounding the supernova.  This low velocity material was and still
is taken as evidence that Sk~-69~202 lived for a time as a red
supergiant, making the (last) transition back to the blue just
shortly, about 20,000 years, before its death.

More successful models considered some alteration to the physics of
convective mixing, the dredge up of helium by rotation for example
(Saio et al. 1988; Weiss et al. 1988; Wang 1991; Langer 1991, 1992),
or altered convective mixing plus low metallicity (Woosley et
al. 1988; Langer et al. 1989; Weiss 1989), or else invoked the effects
of close binary membership on the late evolution (Podsiadlowski \&
Joss 1989; Barkat \& Wheeler 1989; Hillebrandt \& Meyer 1989;
Chevalier \& Soker 1989; de~Loore \& Vanbeveren 1992; Braun \& Langer
1995). The rotational models (e.g., Saio et al.) suffered from an {\sl
ad hoc} prescription for the timing and extent of the mixing process,
essentially substituting one puzzle for another (though, as we shall
see later, they {\sl may} have been essentially correct).  The
``restricted semiconvection" plus low metallicity models (e.g.,
Woosley et al. 1988) also employed several assumptions that have yet
to be proven - that convection follows the Ledoux criterion more
closely than the Schwartzschild and that the LMC is metal poor to the
required extent (a factor of 3 or more in oxygen).  But they were
still regarded as successful until the complicated ring structure,
revealed by Space Telescope, called all single star models into doubt.
The binary star models suffered perhaps from a too great diversity of
possible outcomes and the unavoidable conclusion that SN~1987A was an
unusual event, even in the LMC.

Today, ten years later, we revisit this subject.  Little has been
learned about the (central) supernova itself that changes our views of
the presupernova star.  Its light has continued to decline at a steady
rate and is currently powered by some combination of $^{44}$Ti, a
compact collapsed remnant, and circumstellar shock interaction, just
as was expected. However, the discovery of the complicated double ring
structure and the need to break spherical symmetry prior to the
explosion shows that, at minimum, rotational effects must be included
and, at a maximum, that SN~1987A was a consequence of binary
merger. The latter possibility is discussed in detail elsewhere (and
in this volume) by Podsiadlowski.  Here we will concentrate on single
star models. During the last ten years our understanding of stellar
opacities also has improved.  The new OPAL opacities (Iglesias \&
Rogers, 1996) are considerably different from what was used ten years
ago and, as we shall see, this has important consequences for Sk -69
202.  It has also become possible to include, in a preliminary way,
the effects of rotation on massive stellar evolution and we do so for
Sk -69 202 for the first time here.

We begin with a brief discussion of what causes a star to become a
blue supergiant and then present new calculations of massive star
evolution in the mass range 16 to 22 M\sun. These are non-rotating
models that employ either solar composition or a composition that
might be appropriate to the LMC. The degree of semiconvection is
varied, either reduced throughout the star's life or, for some cases
(Table 2) only after hydrogen burning. There is no obvious need that
the parameters of mixing should be exactly the same on the main
sequence and during advanced burning stages.  Both, red and blue
solutions are obtained and discussed.  Next, we consider the
modification to these models caused by an appreciable amount of
rotation.  Rotation leads to mixing, especially in regions of large
velocity shear like those at the boundaries of convective shells, and
this changes the evolution.  Velocity shear also leads to angular
momentum transport which is followed in the calculations. For the most
part, rotation suppresses the blue solution and makes it harder to
understand SN 1987A.  But in the conclusions, we discuss how
additional angular momentum transport by magnetic fields might lead to
the appreciable dredge up of helium into the envelope.  This would
help make the star blue and also explain large surface abundance
enhancements in nitrogen.

\section{Some generalities}

It was agreed then and still is that the star that exploded, Sk -69
202, a B3 Ia supergiant, had a luminosity of $3 - 6 \times 10^{38}$
erg s$^{-1}$ and a radius of $3 \pm 1 \times 10^{12}$ cm. The
luminosity implies the helium-core mass was 6 $\pm$ 1 M\sun, which if
the effects of mass loss, rotation, and overshoot mixing are ignored,
corresponds to a mass on the main sequence of 18 - 22 M\sun. For a
credible explosion kinetic energy, $1 - 2 \times 10^{51}$ erg,
consistent with the velocity history observed for SN 1987A and the
very early light curve, the star must also have had a hydrogen
envelope mass around 10 M\sun \ in order that the secondary maximum of
the light curve occurs at the right time (e.g., Woosley
1988). Certainly envelope masses below 5 and above 15 M\sun \ are
excluded. Analysis of the neutrino signal gives a mass for the iron
core that collapsed of 1.2 - 1.7 M\sun \ (e.g., Burrows 1988) which is
reasonable for progenitor helium core masses around 6 M\sun. The inner
ring of slow moving nitrogen-rich material alluded to previously also
implies that Sk -69 202 was a red supergiant for some time before
making a transition to a blue star about 20,000 years prior to
exploding. While a red supergiant, the star lost at least 0.045 M\sun
\ (Lundquist \& Fransson 1996) and perhaps as much as several M\sun \ (see
elsewhere in these proceedings). The nitrogen enhancements observed in
the inner ring are N/C $\approx$ 5.0 $\pm$ 2.0 (about 14 times solar)
and N/O $\approx$ 1.1 $\pm$ 0.4 (about 10 times solar, Lundquist \&
Fransson 1996). The outer rings also appear to be nitrogen rich though
less so by a factor of three indicating ejection at an earlier epoch
(Panagia et al. 1996).

In general, increasing the helium core mass makes it harder to get a
blue progenitor. Larger helium cores have larger luminosities to
inflate their hydrogen envelopes. On the other hand a large envelope
mass favors a blue solution. The hydrogen burning shell does not
contribute appreciably to the presupernova luminosity, but its mass
does add to gravity. Making the low density envelope rich in helium
also helps to achieve a blue solution by reducing the opacity and
increasing the mean molecular weight. The envelope will always be
enriched in nitrogen and helium if the star has been a red supergiant
with a deeply convective envelope, but this mixing can be amplified by
rotational effects or perhaps in the common envelope stage of binary
merger. Reducing the metallicity tends to produce a blue supergiant,
both by reducing the envelope opacity and the efficiency of hydrogen
shell burning by the CNO cycle shortly before the supernova.
Obviously, raising the envelope opacity, e.g., because of new
calculations of the relevant atomic physics, also tends to make a red
star. Reduced semiconvection helps make a blue solution, in part by
decreasing the gravitational potential at the hydrogen burning shell
(Lauterborn et al.  1971) and by reducing the carbon-oxygen core mass
developed by a helium core of given size.

One obvious consequence of these rules is that any process that tends to
increase the helium-core mass relative to the hydrogen-envelope mass will
tend to make a red supergiant.  Examples of such processes are mass loss
(short of removing the entire envelope) and rotation.

\section{Computer Models: 16 - 22 M\sun}

We now present a series of model calculations to illustrate the range
of outcomes one might expect for various assumptions regarding the
composition, convection, and rotation. All models in Section~3.1 were
calculated using the KEPLER computer code (Weaver, Zimmerman, \&
Woosley 1978) modified to use, on request, the new OPAL opacities
(Iglesias \& Rogers, 1996) and low-temperature opacities (Alexander
\& Fergusson, 1994).

Where a composition appropriate to the LMC was called for, we used
74.6\% H, 25\% He, 0.052\% C, 0.012\% N, 0.25\%O, 0.056\%Ne, and
0.08\% Fe, all by mass. This is roughly a metallicity of 1/4 solar
consistent with the sum of CNO being (0.30 $\pm$ 0.5)\% in the inner
ring around the supernova (Lundquist \& Fransson 1996). Where used,
{\sl solar} abundances were from Anders \& Grevesse (1989). The
critical reaction rate for $^{12}$C($\alpha,\gamma$)$^{16}$O was taken
equal to the Caughlan \& Fowler (1988) value except where otherwise
noted. Mass loss rates, where included, were from Nieuwenhuijzen et
al. (1990) scaled by a factor (Z$_{\rm surf}$/Z\sun)$^{0.65}$ with
Z$_{\rm surf} \approx 5.6 \times 10^{-4}$ for the LMC models.

\subsection{Non-rotating Models}

Table 1 gives the results for a variety of stellar models. All of
these models used ``restricted semiconvection" in the sense of Woosley
et al.  (1988). That is the mixing criterion, while not quite Ledoux,
did use a small value of semiconvective mixing coefficient, $\alpha$ =
10$^{-4}$ throughout the evolution (here $\alpha$ is a multiplier on
the diffusion coefficient used for transporting composition; D$_{\rm
semi}$ $\approx$ $\alpha$D$_{\rm rad}$). This is to be contrasted with
the standard value used e.g., in Woosley \& Weaver (1995), of 0.1. We
call the latter here ``full semiconvection". If there is an entry for
$\tau_{\rm blue}$, the star died with a surface temperature over
10,000~K and this was the time spent that way following the latest
episode as a RSG. An accompanying entry for $\tau_{\rm red}$ is then
the time spent with surface temperature less than 5,000~K {\sl prior to
this last blue stage} or prior to the SN explosion, if there was no
such blue stage.  The lifetimes may sum to much less than the helium
burning lifetime, about $0.5 - 1 \times 10^6$~yr in all cases, if the
star experienced a blue loop or spent an extended period burning
helium in the blue before becoming red.

We see that using the old opacities - the same as in the code ten years
ago - blue solutions are still obtained for 18, 20, and 22 M\sun \
stars. The 18 M\sun \ model (for zero mass loss) is blue much too long
before the explosion. The inner ring would be much farther out than
observed. In the case of the 20 and 22 M\sun \ models evolved without mass
loss, the final red state just before the last transition to the blue is
short, a feature not emphasized in previous discussions. Each star actually
spends several hundred thousand years as a RSG burning helium, but has an
extended blue loop that ``resets the clock" insofar as Table 1 is concerned.
The short last red phase might eject as much as $\sim$0.1 M\sun \ which
might be enough to explain the inner ring, or at least that position visible
so far, but it is expected that the actual mass in the ring is much larger
(McCray, these proceedings). The star most like the earlier calculations of
Woosley and Weaver is the 22 M\sun \ model that includes mass loss
(de Jager, Nieuwenhuijzen, \& van der Hucht 1985). This loss
suppresses the blue loop and the star spends most of its helium burning
lifetime as a RSG then moves to the blue a proper 20,000 years before
exploding. For the other models the implication of a previous RSG phase with
lots of mass loss followed by a blue wind then a red wind then a blue wind
have yet to be explored, but might be related to the complicated inner and
outer ring structure.

\begin{table} 
\caption{New models --- no rotation}
\label{tbl-1}
\begin{center}\scriptsize
\begin{tabular}{crrrrrrrrrrc}
\tableline
\tableline
\noalign{\vskip 0.1 in}
M & $\dot {\rm M}$ & $\kappa$ & $\tau_{\rm red}$ & $\tau_{\rm blue}$&
M$_{\alpha}$ & M$_{\rm env}$ & L$_{38}$ & He$_s$ & $^{14}$N/$^{12}$C &
$^{14}$N/$^{16}$O & note \\
(M\sun) & & & (ky) & (ky) & (M\sun) & (M\sun) & erg/s & & (\sun) & (\sun)\\
\noalign{\vskip 0.1 in}
\tableline
\noalign{\vskip 0.1 in}
18 & 0   & old  & 190 & 460 & 5.5 & 12.6 & 3.81 & 25 & 2.4 & 1.1 &   \\
18 & 0   & OPAL & 700 &  -  & 5.3 & 12.8 & 3.33 & 27 & 6.4 & 2.6 & a \\
18 & yes & OPAL & 710 &  -  & 5.2 & 11.7 & 3.28 & 27 & 6.8 & 2.7 & a \\
18 & 0   & OPAL & 640 & 25  & 5.5 & 12.6 & 3.91 & 25 & 3.6 & 1.5 & c \\
18 & acc & OPAL & 690 &  -  & 5.3 & 14.9 & 4.05 & 27 & 5.5 & 2.4 & d \\
18 & 0   & OPAL & 680 &  -  & 5.2 & 12.9 & 3.42 & 26 & 4.3 & 1.8 & e \\
18 & yes & OPAL & 710 &  -  & 5.2 & 11.9 & 3.29 & 26 & 5.8 & 2.3 & f \\
18 & 0   & OPAL & 700 &  -  & 5.3 & 12.8 & 3.34 & 25 & 3.8 & 1.6 & g \\
\noalign{\vskip 0.2 in}
20 & 0   & old  & 50  &  30 & 6.5 & 13.6 & 5.12 & 25 & 2.6 & 1.2 & b \\
20 & 0   & OPAL & 600 &  -  & 6.2 & 13.9 & 4.51 & 27 & 6.8 & 2.8 & a \\
20 & yes & OPAL & 580 &  -  & 6.1 & 12.3 & 4.27 & 27 & 6.3 & 2.8 & a \\
20 & 0   & OPAL & 560 &  -  & 6.4 & 13.7 & 4.63 & 26 & 4.1 & 1.8 & a,c \\
\noalign{\vskip 0.2 in}
22 & 0   & old  &  40 &  12 & 7.4 & 14.7 & 6.52 & 25 & 2.8 & 1.3 & b \\
22 & yes & old  & 450 &  20 & 7.2 & 13.6 & 5.95 & 25 & 3.4 & 1.5 & h \\
22 & 0   & OPAL & 540 &  -  & 7.1 & 15.0 & 5.87 & 28 & 7.3 & 3.0 & a \\
22 & yes & OPAL & 520 &  -  & 6.9 & 13.0 & 5.47 & 28 & 7.6 & 3.1 & a \\
\end{tabular}
\end{center}
\tablenotetext{a}{Spends entire post main sequence lifetime in the red.}
\tablenotetext{b}{Extended blue loop after being red since hydrogen
exhaustion.}
\tablenotetext{c}{Metallicity divided by 5 in opacity routine only.}
\tablenotetext{d}{Accretion of 2 M\sun at
$3 \times 10^{-6}$ M\sun \ y$^{-1}$ during
helium burning.}
\tablenotetext{e}{$^{12}$C($\alpha,\gamma)^{16}$O rate times 1.7.}
\tablenotetext{f}{Scale height in mixing length multiplied by 1.5.}
\tablenotetext{g}{Scale height in mixing length multiplied by 2.5.}
\tablenotetext{h}{de Jager, Nieuwenhuijzen, \& van der Hucht (1985) mass
loss rate; see also Chiosi \& Maeder (1986)}
\end{table}

Unfortunately though, the nitrogen enhancements in the outer
atmospheres of all these models are too small to agree with
observations of the circumstellar shell. But at least the stars are
blue. Even this appealing trait is unfortunately lost with the new
OPAL opacities (Table 1). Figure 1 shows the log of the ratio of the
old opacities to the OPAL opacities. A substantial increase in opacity
around 2 10$^5$ K, $\rho \approx 10^7$ g cm$^{-3}$ due to iron has an
important effect on the atmospheric structure. It would be nice to
vary iron separately in the opacity calculation but, as it is, the
opacity tables assume solar ratios and scale according to the total
metallicity. In one case we decreased the metallicity passed to the
opacity routine by an additional factor of 5, but did not change the
abundances used for energy generation. This gave a very nice model at
least with respect to red and blue lifetimes, but we do not believe
the LMC is that metal poor. The calculation only indicates the
sensitivity to the opacity. Another calculation that reduced the
metallicity by only a factor of two (more than the assumed LMC value)
stayed red.

\begin{figure} 
\plotone{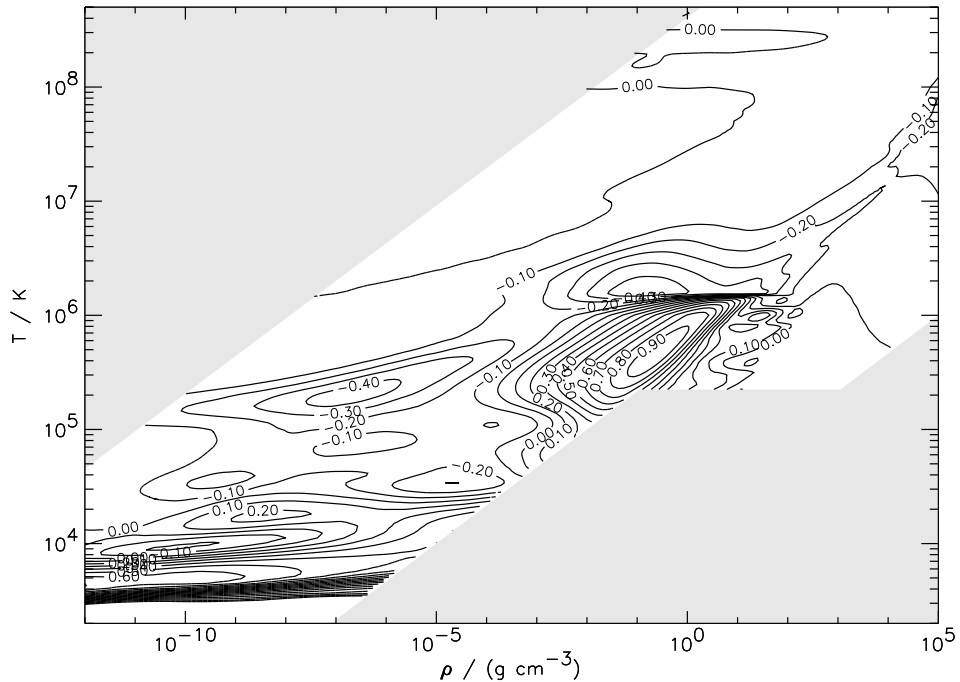}
\caption{The logarithm of the ratio of the old opacities used in
KEPLER to the new OPAL opacities.}
\end{figure}

We also experimented with mass accretion during core helium burning to
investigate the effect of massive hydrogen envelopes.  Interestingly,
a model with a 15 M\sun \ envelope, all that the SN 1987A light curve
would allow, stayed red (cf. Table~1). In our code at least, simply
adding matter to the star does not necessarily make it turn blue. This
constrains some varieties of binary models. We did not experiment with
adding helium rich matter however.

Some binary models (Podsiadlowski, this volume) invoke ``Case C'' mass
transfer after helium burning in order to provoke a common envelope
merger just before the explosion. The timing of this late time
expansion is also governed by the Kelvin-Helmholtz time-scale of the
helium core --- about 20,000 years --- as in the single star case. In
the models we calculated, though, this expansion was always very small
(sometimes zero). The greatest expansion observed was for an 18 M\sun
\ model having solar composition and restricted semiconvection,
$\ltaprx$40\%.  More typically the expansion was $\ltaprx$10\%. The
necessary coincidence that this small expansion led to mass exchange
would make such events rare. However, we have only calculated a
limited number of models and the results are sensitive to the
treatment of convection.

\begin{table} 
\caption{Opal opacities - restricted semiconvection only for $\bar
{\rm A} \ge$ 4}
\label{tbl-2}
\begin{center}\scriptsize
\begin{tabular}{crrrrrrrrc}
\tableline
\tableline
\noalign{\vskip 0.1 in}
M & $\dot {\rm M}$ & comp & $\tau_{\rm red}$ & $\tau_{\rm blue}$&
M$_{\alpha}$ & M$_{\rm env}$ & L$_{38}$ & He$_s$ & note \\
(M\sun) & & & (ky) & (ky) & (M\sun) & (M\sun) & erg/s & (\%) & \\
\noalign{\vskip 0.1 in}
\tableline
\noalign{\vskip 0.1 in}
16 & 0   & LMC  &  60 &  95 & 4.2 & 11.9 & 2.72 & 25  &   \\
16 & yes & LMC  & 100 &  85 & 4.2 & 11.5 & 2.61 & 25  &   \\
\noalign{\vskip 0.2 in}
18 & 0   & LMC  &  -  &  80 & 4.8 & 13.3 & 3.96 & 25  & a \\
18 & yes & LMC  & 100 &  35 & 4.8 & 12.7 & 3.79 & 25  &   \\
18 & 2x  & LMC  &  75 &  45 & 4.7 & 12.4 & 3.59 & 25  &   \\
18 & 0   & LMC  &  -  &   - & 5.0 & 13.1 & 4.16 & 25  & b \\
18 & yes & solar& 450 &  -  & 4.7 & 11.3 & 3.76 & 36  & c \\
18 & yes & solar&1000 &  -  & 5.2 &  8.7 & 4.80 & 37  & c,d \\
\noalign{\vskip 0.2 in}
19 & yes & LMC  &  -  &  50 & 5.3 & 13.3 & 4.37 & 25  & a  \\
\noalign{\vskip 0.2 in}
20 & 0   & LMC  &  30 &  20 & 5.7 & 14.4 & 5.40 & 25  &   \\
20 & yes & LMC  &  50 &   7 & 5.7 & 13.7 & 5.15 & 26  &   \\
\end{tabular}
\end{center}
\tablenotetext{a}{Surface T drops below 10,000 K, but the star moves back to
blue very quickly} 
\tablenotetext{b}{Old opacities. Surface T has a
dip at the end of central He burning but never drops below 10,000 K.}
\tablenotetext{c}{Star does not get blue before supernova occurs}
\tablenotetext{d}{Semiconvective mixing parameter 0.1 rather than
10$^{-4}$ even after hydrogen burning.}  
\end{table}

\begin{figure} 
\plotfiddle{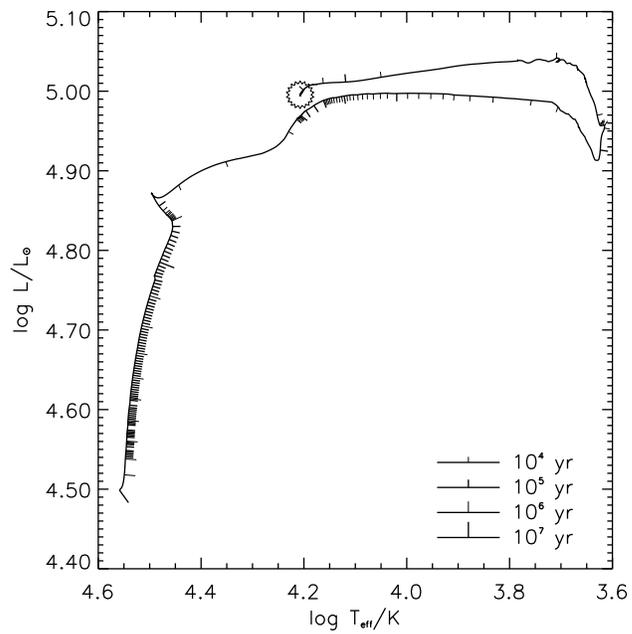}{8 cm}{0}{50}{50}{-150}{-90}
\caption{The HR-diagram for an 18 M\sun \ star evolved using OPAL
opacities and restricted semiconvection only for $\bar{\rm A} \ge$ 4.
The toothed circle marks the position in the HRD where the star would
explode as SN.  It spends its last $\approx$35,000~yr as a blue
supergiant.}
\end{figure}

\begin{figure} 
\plotfiddle{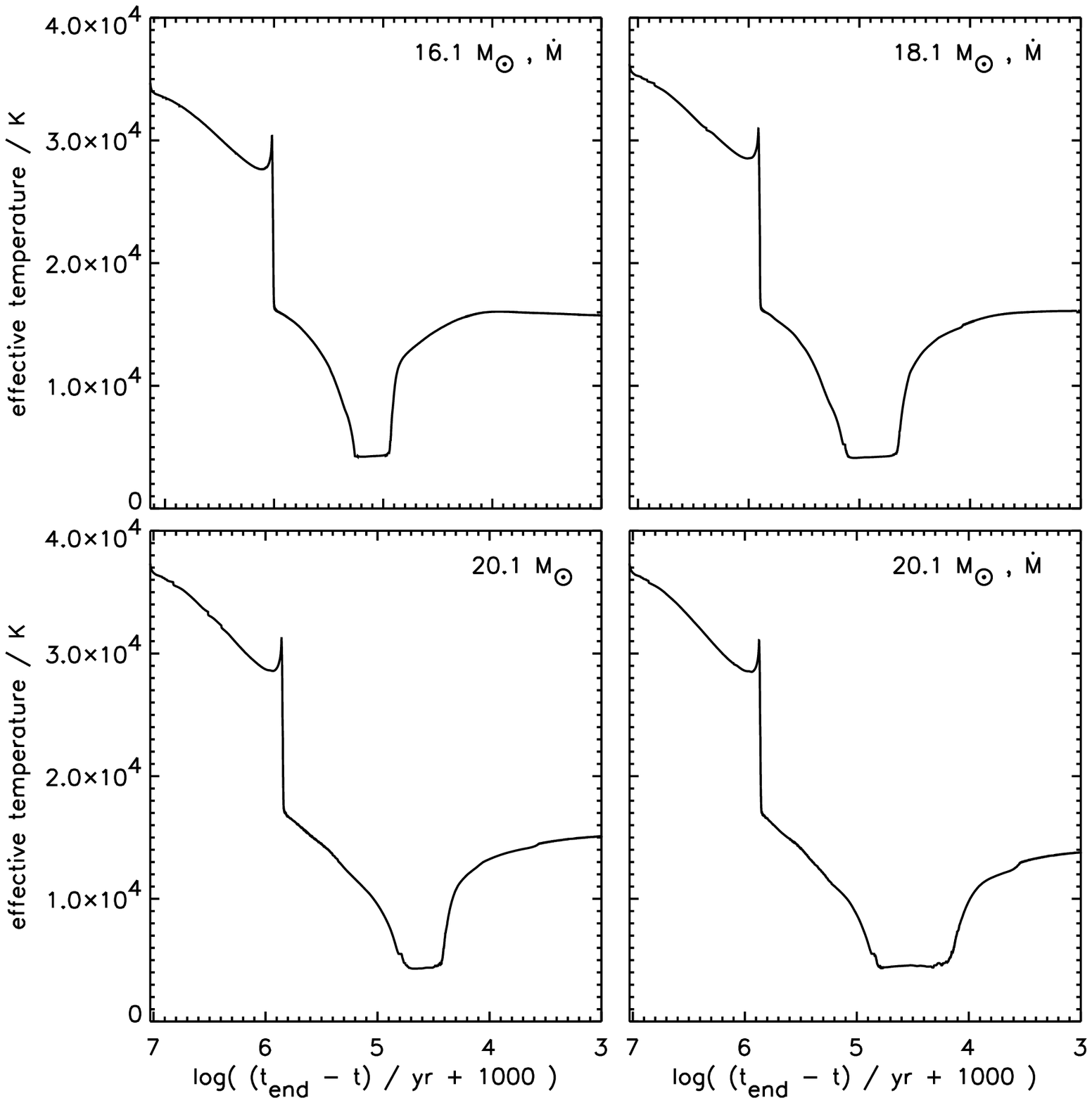}{8 cm}{0}{50}{50}{-150}{-80}
\caption{Effective temperature as a function of time for four models
with blue SN progenitors.  They are calculated starting with LMC
composition and using OPAL opacities, restricted semiconvection only
for $\bar{\rm A} \ge$ 4 (full semiconvection otherwise) and mass loss
(see text) where indicated by ``$\dot{\rm M}$''.  The time-scale of
the blue phase before the SN ranges from about 10,000~yr (20~M\sun \
with mass loss) to almost 100,000~yr (16 M\sun).}
\end{figure}

We also experimented with the convection prescription. The uncertainty
regarding semiconvection is such that there is no compelling reason
that the multiplier on the diffusion coefficient should be the same on
the main sequence and during advanced stages. We calculated a series
of models in which reduced semiconvection was only employed in regions
where the mean atomic weight $\bar {\rm A} \ge$ 4. The results were
appreciably different from the $\alpha$ = constant cases. Table 2
summarizes the results.

Even using the OPAL opacities there are now a number of blue solutions
and some of these, e.g., the 18 M\sun \ models, are red for a long
time moving to the blue only at the end. The low surface abundance
enhancements of nitrogen remain of some concern however. Only the
solar metallicity models had sufficiently deep convective envelopes to
give large values, as indicated by the large helium abundances in the
table.

\begin{table} 
\caption{Nucleosynthesis and remnant masses}
\label{tbl-3}
\begin{center}\scriptsize
\begin{tabular}{rrrrrrr}
\tableline
\tableline
\noalign{\vskip 0.1 in}
Element & LMC-18 & SOL-18 & LMC-20 & SOL-20 & LMC-22 & SOL-22 \\
\noalign{\vskip 0.1 in}
\tableline
\noalign{\vskip 0.1 in}
Fe core & 1.27 & 1.42 & 1.62 & 1.74 & 1.36 & 1.82 \\
\noalign{\vskip 0.1 in}
Remnant & - & 1.76 & - & 2.06 & -  & 2.02 \\
\noalign{\vskip 0.2 in}
H    & 8.71 & 7.89 & 9.31 & 8.24 &  9.91 & 8.79 \\
He   & 7.02 & 6.28 & 8.07 & 6.72 &  8.82 & 7.51 \\
C    & 0.19 & 0.25 & 0.27 & 0.21 &  0.34 & 0.24 \\
N    & 0.015 & 0.057 & 0.017 & 0.060 & 0.019 & 0.067 \\
O    & 0.28 & 1.13 & 0.66 & 1.94 & 1.02 & 2.38 \\
Ne   & 0.27 & 0.28 & 0.10 & 0.11 & 0.063 & 0.07 \\
Mg   & 0.058 & 0.055 & 0.026 & 0.031 & 0.047 & 0.042 \\
Si   & -    & 0.14 &  - & 0.29 & - & 0.36 \\
S    & -    & 0.055 & - & 0.15 & - & 0.17 \\
Ar   & -    & 0.0087 & - & 0.026 & - & 0.028 \\
Ca   & -    & 0.0068 & - & 0.014 & - & 0.017 \\
$^{56}$Ni & - & 0.066 & - & 0.088 & - & 0.205 \\
\end{tabular}
\end{center}
\end{table}

The nucleosynthesis in Table 3 is taken from the 18, 20, and 22~M\sun
\ models that used old opacities, no mass loss, and no rotation. They
were the only models evolved so far to the presupernova state. The
solar metallicity versions of these models used the large
semiconvection parameter and are taken from Woosley \& Weaver
(1995). Oxygen is an interesting diagnostic of the models, a value
appreciably less than 1.0 M\sun \ being strongly suggestive of
restricted semiconvection.

All numbers in Table 3 are in solar masses.  All ``SOL'' stars ended
their lives as {\sl red} supergiants. The ``LMC'' models used the LMC
composition, old opacities, and small semiconvection parameter. They all
ended their lives as {\sl blue} supergiants.

\subsection{Rotating Models}

As the complicated ring structure around SN 1987A indicates, spherical
symmetry has been broken. If the star was not a binary, rotation must
have played a role. Rotation can also enhance deep mixing and that
might help to enrich the atmosphere in helium and nitrogen (Langer 1991).

A recent analytic model by Meyer (1997) suggests that the outer rings
of SN~1987A are formed as a result of an aspheric slow RSG wind which
has been heated and ionized during the pre-SN blue phase of
SK~-69~202.  Whether a single star carries enough angular momentum to
form the density contrast between equator and pole (at least $\sim$ 4)
needed in Meyer's model is not yet clear but might not be excluded
(see e.g. Asida \& Tuchman 1995).

\begin{table}  
\caption{Models that include rotation}
\label{tbl-4}
\begin{center}\scriptsize
\begin{tabular}{crrrrrrrrrrr}
\tableline
\tableline
\noalign{\vskip 0.1 in}
M & $\dot {\rm M}$ & comp & v$_{\rm rot}$ & $\tau_{\rm red}$ 
& $\tau_{\rm blue}$ & M$_{\alpha}$ & M$_{\rm env}$ & L$_{38}$ & He$_s$ & 
$^{14}$N/$^{16}$O & $^{14}$N/$^{12}$C \\
(M\sun) & & & km/s & (ky) & (ky) & (M\sun) & (M\sun) & erg/s & (\%) & (\sun) &
(\sun)\\
\noalign{\vskip 0.1 in}
\tableline
\noalign{\vskip 0.1 in}
18 & yes  & LMC  & 0   & 615 &  -  & 5.0 & 12.0 & 3.24 & 28  &  6.3 &  4.2 \\
18 & yes  & LMC  & 20  & 240 &  -  & 5.8 & 11.0 & 5.36 & 39  &  7.9 & 11.3 \\
18 & yes  & LMC  & 200 & 715 &  -  & 7.2 &  8.0 & 7.77 & 46  & 13.6 & 25.1 \\
\end{tabular}
\end{center}
\end{table}

A series of rotating models were calculated using the G\"ottingen
stellar evolution code (Table~4). This had a similar, though not
identical treatment of semiconvection to the ``restricted'' cases done
with KEPLER and a similar treatment of mass loss (cf. Langer et al. 1989).
The assumed starting
model was rigidly rotating on the main sequence with an equatorial
velocity given in Table 4. Angular momentum transport was treated as
described in Langer et al. (1997; see also Endal \& Sofia 1976, 1978,
1981; Pinsonneault et al. 1989) and included terms for convection,
semiconvection, Eddington Sweet circulation, the Solberg-Hoiland
instability, as well as secular and dynamic shear instability.
Although the transport of angular momentum is treated in diffusion
approximation, convective regions tended to rotate rigidly, because
the convective turnover time-scale is usually small in comparison to
the live-time of convective regions and the time-scale on which the
convective region changes its properties, e.g. its density structure
and total angular momentum.

\begin{figure} 
\plotone{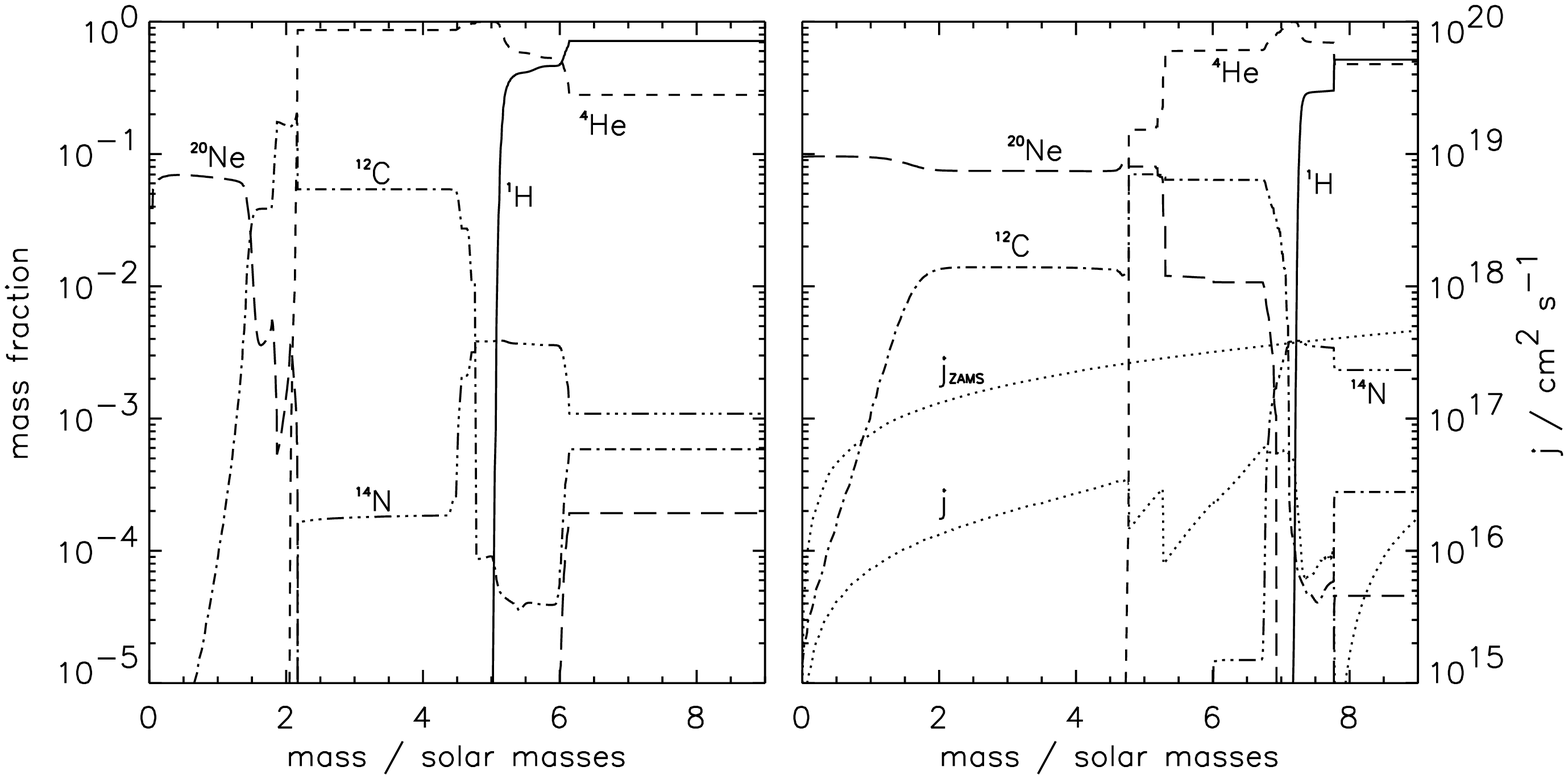}
\caption{Composition and specific angular momentum of two LMC 18 M\sun
\ stars evolved to carbon depletion with (right, last line in Table 4)
and without (left, first line in Table 4) rotation.  The rotating
model had a ZAMS equatorial rotation velocity of 200 km/s at the
surface and was rigidly rotating (upper dotted line, indicated by
``j$_{\rm ZAMS}$'').  The final distribution of j pictured is a result
of angular momentum loss of the star due to mass loss from the surface
and redistribution of the angular momentum due to convetion and
rotational instabilities.  The convective envelope of the RSG, which
reaches down to about 7.8~M\sun, is chemically homogeneous and
rigidly rotating, i.e. the major part of the angular momentum of the
star is located close to the surface of the star.}
\end{figure}

Sample model results at carbon depletion are given in Table 4 for two
rotating 18~M\sun \ stars (the zero rotation case is also given for
comparison). Both rotating stars developed deep convective envelopes
and had large helium and nitrogen enhancements. However, both ended up
as red supergiants. The main culprit here is the growth of the helium
core (Figure 4) due to additional rotationally induced mixing (Langer
et al. 1997). A larger helium core implies a bigger luminosity which
is bad enough, but the larger luminosity also causes more mass
loss. Both effects make a blue solution difficult.


\section {Speculations and Conclusions}

Our conclusions here are mostly negative - without incredibly large
decreases in metallicity, the new opacities preclude an acceptable
blue solution for Sk -69 202 even if the old treatment of restricted
semiconvection is used. Using restricted semiconvection only after
hydrogen burning gives more acceptable results, but the surface
abundances of N and He remain problematic. A moderate amount of
rotation and a reasonable - but poorly understood - prescription for
angular momentum transport gives the desired deep mixing, but too
large helium cores to make a blue progenitor.

Before surrendering the possibility of a single star solution,
however, and accepting that SN~1987A was a very improbable event, we
consider another possibility - that rotation was the cause of the ring
structure, the abundance anomalies in the rings, and the blue
progenitor.

Our rotating presupernova stars (Fig.~4) end up with very large
velocity shears at the outer edge of the helium core. Inclusion of
magnetic fields, or other mixing mechanisms not considered so far
(Acheson 1978, Spruit \& Phinney 1997), might lead to additional
helium being dredged up into the envelope.  Some kind of additional
angular momentum transport is indicated because the continued
evolution of these stars, using the same angular momentum transport
scheme, ends up making iron cores with 100 times the specific angular
momentum of the Crab pulsar (Heger, Woosley, \& Langer 1997). While we
can provide no models at the present time, any mechanism that reduced
the helium core while simultaneously increasing helium in the envelope
would tend strongly towards a blue solution.  Whether this would
naturally occur 20,000 years prior to the explosion remains to be
demonstrated (see also, e.g., Saio et al. 1988).

A rotating RSG with a deeply convective envelope that moved back to
the blue would also concentrate a large fraction of the angular
momentum in the outer few hundredths of a solar mass (Heger \& Langer
1997). This would lead to near Keplerian motion in the outer layers
and asymmetry in the mass loss.  Of special interest is the {\sl
first} time this occurs for stars which experience blue loops, as many
of our models do.  Whether this asymmetry could be enough to explain
the complex outer ring structure is unknown, at least it might be an
explanation for the inner ring, but we consider the single star models
for SN~1987A worthy of continued study.

\acknowledgments

We are grateful to F. Meyer and H.C. Spruit for delighting
discussions.  This work has been supported by the National Science
Foundation (AST -94 17171) and NASA (NAG5 3434). A. Heger was supported
by a ``DAAD-\-Dok\-to\-ran\-den\-sti\-pen\-di\-um aus den Mit\-teln
des 2. Hoch\-schul\-pro\-gramms''
while in Santa Cruz.

\end{document}